\documentclass[manuscript]{acmart}
\def\BibTeX{{%
    \normalfont B\kern-0.5em{\scshape i\kern-0.25em b}\kern-0.8em\TeX}}

\copyrightyear{2020} 
\acmYear{2020} 
\setcopyright{acmcopyright}\acmConference[RecSys '20]{Fourteenth ACM Conference on Recommender Systems}{September 22--26, 2020}{Virtual Event, Brazil}
\acmBooktitle{Fourteenth ACM Conference on Recommender Systems (RecSys '20), September 22--26, 2020, Virtual Event, Brazil}
\acmPrice{15.00}
\acmDOI{10.1145/3383313.3412224}
\acmISBN{978-1-4503-7583-2/20/09}

\begin{document}

\title{Improving One-class Recommendation with Multi-tasking on Various Preference Intensities}

\author{Chu-Jen Shao}
\email{r07922044@ntu.edu.tw}
\affiliation{%
  \institution{National Taiwan University}
  \country{Taiwan}
}

\author{Hao-Ming Fu}
\email{r06922092@ntu.edu.tw}
\affiliation{%
  \institution{National Taiwan University}
  \country{Taiwan}
}

\author{Pu-Jen Cheng}
\email{pjcheng@csie.ntu.edu.tw}
\affiliation{%
  \institution{National Taiwan University}
  \country{Taiwan}
}

% \affiliation{%

% \affiliation{%

% \affiliation{%

% \affiliation{%

% \affiliation{%




\begin{abstract}
  In the one-class recommendation problem, it’s required to make recommendations basing on users’ implicit feedback, which is inferred from their action and inaction. Existing works obtain representations of users and items by encoding positive and negative interactions observed from training data. However, these efforts assume that all positive signals from implicit feedback reflect a fixed preference intensity, which is not realistic. Consequently, representations learned with these methods usually fail to capture informative entity features that reflect various preference intensities.
  
  In this paper, we propose a multi-tasking framework taking various preference intensities of each signal from implicit feedback into consideration. Representations of entities are required to satisfy the objective of each subtask simultaneously, making them more robust and generalizable. Furthermore, we incorporate attentive graph convolutional layers to explore high-order relationships in the user-item bipartite graph and dynamically capture the latent tendencies of users toward the items they interact with. Experimental results show that our method performs better than state-of-the-art methods by a large margin on three large-scale real-world benchmark datasets.
\end{abstract}


\ccsdesc[500]{Information systems~Recommender systems}

\keywords{collaborative filtering; one-class recommendation; graph convolutional network; implicit feedback}

\maketitle
\vspace{-0.3cm}
\section{Introduction}
User feedback in recommendation systems can be divided into two types: implicit feedback, such as clicking or setting as bookmarks, and explicit feedback, such as rating a film with 1-5 stars. In general, implicit feedback is easier to obtain than explicit feedback. Thus, making recommendations with only implicit feedback is indispensable. This type of problems are referred to as one-class recommendation \cite{occf}.

There are several efforts proposed to solve one-class recommendation problems. For example, model-based methods \cite{BPR, NCF} aim to learn a vector representation for each user and item and apply some kernel, such as inner product for matrix factorization (MF) \cite{MF}, to measure similarity. On the other hand, graph-based methods \cite{He7} construct a user-item bipartite graph from historical interactions and utilize random walk on it to explore user interests and make recommendations. In recent years, hybrid approaches \cite{Yang8, NGCF9} combining model-based and graph-based methods have been developed. They explore high-order relationships on the bipartite graph and encode this information into learned entity representations, resulting in remarkable improvements in one-class recommendation tasks.

However, there are some weaknesses in existing works. First, users may have strong or weak preferences toward their interacted items while there is no difference between them when being observed in implicit feedback. Existing methods generally assume that these positive interactions reflects a fixed preference intensity, which might be unrealistic. Besides this, we observe that the representations learned from training samples do not effectively generalize to validating samples. Specifically, it's observed that the model predicts high preference only for items that a user interacted with in training samples, but not those interacted in validating cases.

We address these two issues altogether by introducing a multi-tasking framework that takes variable preference intensities into consideration. In this framework, the attentive graph convolutional layer is designed to enhance the preference intensity of observed interactions and explore higher-order relationships. Representations in different levels of preference intensities are required to satisfy their corresponding objectives. Consequently, we argue that the proposed multi-tasking framework greatly improves the expressiveness and robustness of learned representation. Offline experiments show that our proposed method outperforms existing approaches on datasets of various size and sparsity on preference ranking task for recommendation systems.

\begin{figure*}
  \centering
  \includegraphics[width=\linewidth]{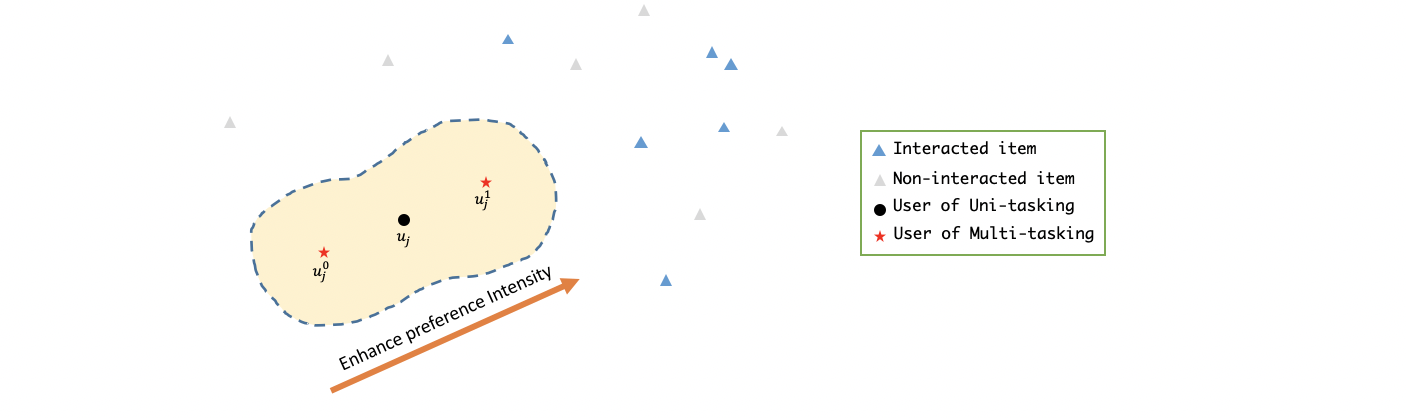}
  \caption{Multi-tasking on various preference intensities in learned representation space. We learn multiple vector representation for $u_j$. They lie in the bar-shaped area, indicating different preference intensities.}
  \label{fig:peanut}
  \Description{An illustration of the effect of our approach in learned embedding space.}
\end{figure*}

\section{Related Works}

% \subsection{Model-based collaborative filtering(CF)}
\textbf{Model-based Collaborative Filtering} approaches aim to capture the relationships between users and items by learning embedding vectors to fit the interaction data. For example, Matrix Factorization \cite{MF} maps users and items into a shared latent vector space by decomposing the user-item interaction matrix. What’s more, to handle implicit data that were not observed in training samples, Bayesian Personalized Ranking (BPR) \cite{BPR} and K-order \cite{K-order} incorporate ranking losses to focus on relative preference rather than the absolute rating score. Recently, deep neural network models have been largely adopted to model the nonlinear relationships between users and items. For instance, NeuMF \cite{NCF} and LRML \cite{Yi5} successfully introduce neural structures into model-based CF approaches.

% \subsection{Graph-based Collaborative Filtering}
\textbf{Graph-based Collaborative Filtering} is another series of methods that makes use of the user-item bipartite graph constructed by users, items and observed interactions between them. If entities are close or reachable through paths in the graph, they are expected to have stronger preferences or similarities. Label propagation and random walk are often used approaches to explore such relations. \cite{Gori6}, \cite{He7}, \cite{Yang8} are some examples taking this route.

% \subsection{Graph Convolutional Network (GCN)}
\textbf{Graph Convolutional Network} can be viewed as a hybrid approach that explores high-order neighboring relations on the graph to learn representation for entities. It inherits the advantages of both model-based and graph-based approaches and achieves better performances. In recent years, a series of GCN-based works have been proposed, GC-MC \cite{Kipf10} employs one graph convolutional layer to exploit the first-order relationships, PinSage \cite{PinSage11} incorporates multiple graph convolutional layers and neighbor sampling strategy for Pinterest image recommendation. NGCF \cite{NGCF9} combines information from different orders of neighbors to form more informative representation.

% Our work leverages the graph convolutional 
% network and further improves it by involving an attention mechanism. Besides, we design a multi-tasking framework considering various preference intensities and make the learned representations more robust and expressive.

\section{Proposed Model}

\subsection{Multi-tasking on Various Preference Intensities}

\begin{figure*}[t]
  \centering
  \includegraphics[width=\linewidth]{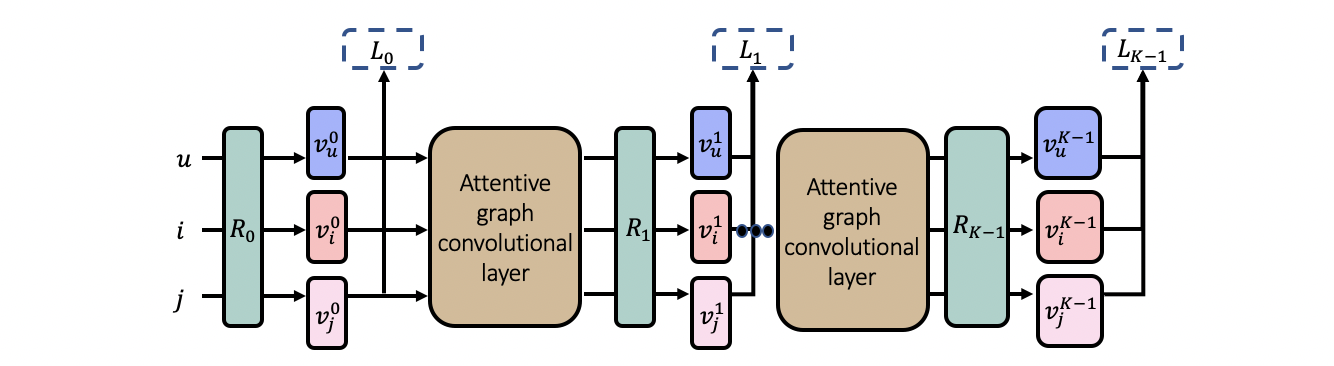}
  \caption{An overview of our model. We establish a lookup table to learn user and item vectors in $R_0$. Each attentive graph convolutional layer following it obtains a new set of representation $R_i$ that assumes stronger preference intensities. We calculate the BPR loss of each $R_i$ and average them to form the total loss. This requires each $R_i$ to satisfy the BPR loss constraints.}
  \label{fig:model}
  \Description{An overview of our model.}
\end{figure*}

We transform the representation of an entity via the graph convolution operation, which calculates the weighted sum of its own representation vector and that of its neighbors on the user-item bipartite graph. We assume that the graph convolution operation augments the bonds between an entity and its neighbors and thus enhances the preference intensity of these user-item pairs. To consider various preference intensities, we design a multi-tasking framework, where each sub-task represents a certain level of preference intensity. Fig. \ref{fig:peanut} shows the conceptualization of our proposed method. We require our learned representation to be robust in a bar-shaped region instead of only at a spot where a single vector locates. For a user, the orientation of the bar-shaped region is along the axis that points to the cluster of interacted items, reflecting stronger (closer to items) and weaker (farer from items) preference intensities. Such determination of orientation is obtained from the fact that in reality, an interaction can infer stronger or weaker preference. We achieve such effect by multi-tasking, learning multiple representation for an entity that locates on different part of the bar-shaped region. In our case, each of the representations corresponds to different preference intensity assumption, guaranteeing the robustness of learned representation against preference intensity variance. To acquire embedding vectors corresponding to different preference intensities, we adopt attentive graph convolution repetitively. This way, the resulting representations lie along the bar-shaped region, each required to satisfy its own objective term.

\subsection{Model Overview}

Fig. \ref{fig:model} is an overview of the model we propose. We recommend items to users by predicting a preference score for each user-item pair. During evaluation, we rank all items for a user from the highest preference score to the lowest and assess the quality of the ranking. To obtain preference scores of each user-item pair, our model learns a vector representation for each user and item in the dataset. The preference score of a user-item pair is then represented by the inner product of the user vector and the item vector. We apply multi-tasking, $K$ tasks for instance in the example below, to learn more robust representation vectors. Specifically, we build a trainable lookup table ${R^0}$. With $K-1$ attentive graph convolutional layers following ${R^0}$, we obtain the transformed representations ${R^1}$, …, ${R^{K-1}}$, respectively. Note that ${R^{l}} \in \mathbb{R}^{(m+n) \times {d_{l}}}$ is the set of representations of entities in the $l_{th}$ layer, where $d_{l}$ denotes the dimension of the representation and $m$, $n$ represent the sizes of user set $U$ and item set $I$, respectively. The details of the transformation is illustrated in Section \ref{sub:3_3}.

Following this, we apply the BPR loss, which is popular for ranking tasks for its promising performance, on each of the representation sets. This provides us with loss terms corresponding to each representation set, denoted as ${L_0}$, ${L_1}$, …, ${L_{K-1}}$. Finally, we average the loss of each task to form the total loss, which is minimized to train the model in an end-to-end manner.

% \begin{figure}[h]
%   \centering
%   \includegraphics[width=\linewidth]{GCN_fig_ver1.png}
%   \caption{Attentive graph convolutional layer. We obtain a new vector $e^{l+1}_{user_j}$ in layer $l+1$ from user and item vectors in layer $l$. Note that $\oplus$ denotes element-wise sum and $att_i$ is automatically learned by an attention network $f()$.}
%   \Description{The details of attentive graph convolutional layer.}
% \end{figure}

% \vspace{-0.5cm}
\begin{table}
  \caption{Statistics of three datasets.}
  \label{tab:dataset}
  \begin{tabular}{c|ccc}
    \toprule
    Dataset&Gowalla&Amazon-book&Yelp2018\\
    \midrule
    Users ($\lvert U \rvert$) & 29858 & 52643 & 45919\\
    Items ($\lvert I \rvert$) & 49081 & 91599 & 45538\\
    Interactions & 1027370 & 2984108 & 1185065\\
    Density & 0.084\% & 0.062\% & 0.056\%\\
  \bottomrule
\end{tabular}
\end{table}

\begin{table*}[t]
  \caption{Overall performance. \% Improv. denotes the percentage of improvement with respect to the underlined best baseline method in each dataset. The best performance of each column is shown in bold numbers.}
  \label{tab:perf}
  \begin{tabular}{c|c|c|c|c|c|c}
    \toprule
    &\multicolumn{2}{c|}{Gowalla} &\multicolumn{2}{c|}{Amazon-book}
    &\multicolumn{2}{c}{Yelp2018}\\
    &  Recall@20 & NDCG@20 & Recall@20 & NDCG@20 & Recall@20 & NDCG@20\\
    \midrule
    MF-BPR & 0.1455 & 0.2204 & 0.0270 & 0.0558 & 0.0550 & 0.0703 \\
    HOP-Rec & \underline{0.1575} & \underline{0.2303} & 0.0285 & 0.0572 & \underline{0.0654} & \underline{0.0832} \\
    GC-MC & 0.1475 & 0.2114 & 0.0313 & 0.0594 & 0.0639 & 0.0806 \\
    PinSage & 0.1472 & 0.2155 & 0.0302 & 0.0606 & 0.0641 & 0.0814 \\
    NGCF & 0.1547 & 0.2237 & \underline{0.0344} & \underline{0.0630} & 0.0634 & 0.0799 \\
    Ours & \textbf{0.1664} & \textbf{0.2357} & \textbf{0.0378} & \textbf{0.0673} & \textbf{0.0696} & \textbf{0.0867} \\
  \midrule
    \% Improv. & +5.65\% & +2.34\% & +9.88\% & +6.83\% & +6.42\% & +4.21\% \\
  \midrule
  \bottomrule
\end{tabular}
\end{table*}

\begin{table*}[t]
  \caption{Results of applying our multi-tasking framework on baseline models. Notice that our model adopts both multi-tasking and attention mechanism. The best performance of each column is shown in bold numbers.}
  \label{tab:att}
  \begin{tabular}{c|c|c|c|c|c|c}
    \toprule
    &\multicolumn{2}{c|}{Gowalla} &\multicolumn{2}{c|}{Amazon-book}
    &\multicolumn{2}{c}{Yelp2018}\\
    & Recall@20 & NDCG@20 & Recall@20 & NDCG@20 & Recall@20 & NDCG@20\\
    \midrule
    GC-MC$_{multi-task}$ & 0.1560 & 0.2301 & 0.0333 & 0.0624 & 0.0651 & 0.0818 \\
    PinSage$_{multi-task}$ & 0.1570 & 0.2290 & 0.0317 & 0.0607 & 0.0649 & 0.0812 \\
    NGCF$_{multi-task}$ & 0.1586 & 0.2287 & 0.0341 & 0.0632 & 0.0654 & 0.0809 \\
    Ours & \textbf{0.1664} & \textbf{0.2357} & \textbf{0.0378} & \textbf{0.0673} & \textbf{0.0696} & \textbf{0.0867} \\
  \bottomrule
  \end{tabular}
\end{table*}

\begin{table*}[t]
  \caption{Effect of including different number of tasks in multi-tasking. Resource limitation hinders us from using K=3, 4 in the Amazon-book dataset. They are shown as -. The best performance of each column is shown in bold numbers.}
  \label{tab:taskL}
  \begin{tabular}{c|c|c|c|c|c|c}
    \toprule
    &\multicolumn{2}{c|}{Gowalla} &\multicolumn{2}{c|}{Amazon-book}
    &\multicolumn{2}{c}{Yelp2018}\\
    Task numbers (K) & Recall@20 & NDCG@20 & Recall@20 & NDCG@20 & Recall@20 & NDCG@20\\
    \midrule
    1 ($R_0$) & 0.1455 & 0.2204 & 0.0270 & 0.0558 & 0.0550 & 0.0703 \\
    2 ($R_0$ + $R_1$)& \textbf{0.1664} & \textbf{0.2357} & \textbf{0.0378} & \textbf{0.0673} & \textbf{0.0696} & \textbf{0.0867} \\
    3 ($R_0$ + $R_1$ + $R_2$)& 0.1612 & 0.2297 &    -   &    -   & 0.0674 & 0.0833 \\
    4 ($R_0$ + $R_1$ + $R_2$ + $R_3$)& 0.1505 & 0.2214 & - & - & 0.0643 & 0.0814 \\
  \bottomrule
\end{tabular}
\end{table*}

\subsection{Attentive Graph Convolutional Layer}
\label{sub:3_3}

% \begin{figure}[h]
%   \centering
%   \includegraphics[width=\linewidth]{attention_fig_ver1.png}
%   \caption{Attention mechanism with attention model $A$. For $user_j$, we feed concatenated $e^{l}_{user_j}$ and vectors of each of its neighboring items $e^{l}_{item_{nbr_i}}$ (as well as $e^{l}_{user_j}$ itself) into the attention network $A$. $A$ then outputs the logit of each user-item pair. These logits go through a softmax function and obtain the attention weights $att_i$, which attentive convolutional layers use to calculate $v^{l}_{ws_{user_j}}$. The structure of the attention network $A$ we use is a 2-layer MLP with 128, 1 nodes respectively.
% }
%   \Description{Attention mechanism with attention model $A$.}
% \end{figure}

We first construct the user-item bipartite graph $\mathcal{G}=(\mathcal{V}, \mathcal{E})$ from the implicit feedback, where the vertex set ${\mathcal{V}}=U \cup I$ and the edge set $\mathcal{E}$ is composed of every positive interactions in the training set. We design the attentive graph convolutional layer that transforms a set of representation into another one. For example, the $l_{th}$ attentive graph convolutional layer takes ${R^{l-1}}$ and the bipartite graph $\mathcal{G}$ as input and transforms it into ${R^{l}}$.

The desired representation of an entity $e \in \mathcal{V}$ in the $l_{th}$ layer is denoted as $v^l_e$. It is obtained by element-wisely weighted sum the ${l-1}_{th}$ layer representations of itself and its neighbors, followed by a linear transformation and a non-linear activation function. The process can be formulated as follows:

    \begin{center}
    $v^l_e = \sigma({\sum\limits_{j \in {N(e)\cup \{e\}}} a^l_{ej} v^{l-1}_jW^l} )$,
    \end{center}
where $\sigma(.)$ is the activation funciton, $N(e) \subset \mathcal{V}$ represents the set of neighbors of $e$ in $\mathcal{G}$, $W^l \in \mathbb{R}^{{d_l-1} \times {d_l}}$ denotes the learnable transformation matrix in the $l_{th}$ attention graph convolutional layer and $a^l_{ej} \in \mathbb{R}$ represents the learned attentive weight in the $l_{th}$ layer between $e$ and its neighbor $j$.

In order to capture the importance between an entity and its neighbors dynamically, we allow our model to learn the weights automatically by using an attention network f(.). f(.) is a two-layer MLP which takes the concatenation of the entity vector and its neighbor vector (one neighbor a time) as input and outputs a single logit for each of them. For an entity being considered, the logits of its neighbors (and itself) go through a softmax function to form attentive weights $a^l_{ej} \in \mathbb{R}$. We can formulate this procedure as:

    \begin{center}
    $a^l_{ej} = \frac{exp(f(v^{l-1}_e, v^{l-1}_ j))}{\sum\limits_{k \in {N(e)\cup \{e\}}}exp(f(v^{l-1}_e, v^{l-1}_k))}$,
    
    and $f(x, y) = \sigma(W^l_{att_2}\sigma(W^l_{att_1} (x||y) + b^l_1) + b^l_2)$
    \end{center}, 
where || denotes concatenation of two vectors and $W^l_{att_1}, W^l_{att_2}$ are learnable matrices in the attention network. Notice that the representation of all entities, $v^l_e$, form the representation set of the $l_{th}$ layer, referred to as $R^l$.

\subsection{Traning and Evaluation}
Given a triplet ${\{u, i, j\}}$ where user $u$ has interacted with item $i$ and not interacted with item $j$, the BPR loss is designed to maximize the difference of preference scores between the positive pair $(u, i)$ and the negative pair $(u, j)$.
Following this idea, we define K loss terms, ${L_0}$, ${L_1}$, …, ${L_{K-1}}$ corresponding to the K sub-tasks as follows:
    \begin{center}
    ${L_l = -\sum\limits_{\{u,i,j\} \in D}\ln{\sigma(v^l_u \cdot v^l_i - v^l_u \cdot v^l_j)}}$,
    \end{center}
where $D$ denotes the entire training set and $\sigma$ is the sigmoid function.

% Suppose that we are given a set of representation R containing user vectors vuj and item vectors vik. A single input sample of our model is a triplet: user j, item k, item l, where user j interacted with item k but not item l. Given such information, we aim to assign a higher preference score to the pair of user j and item k than to user j and item l. We use cross entropy, namely, sigmoid function followed by logarithm, to maximize such difference:

With the BPR loss terms of each acquired representation set calculated as ${L_0}$, ${L_1}$, …, ${L_{K-1}}$, we simply use their mean as our overall total loss:
    \begin{center}
    $L_{total} = \frac{1}{K}\sum_{l=0}^{k-1}{L_l}$
    \end{center}
$L_{total}$ is then minimized by the Adam optimizer to update the variables in the model in an end-to-end manner.

Under our framework, each set of representation is updated by gradients from multiple tasks. To be precise, ${R^l}$ obtains gradients from the loss terms of ${R^l}$, ${R^{l+1}}$, …, $R^{K-1}$, where $R^l$ is the last set of representation. To consider various preference intensities and improve model robustness, we concatenate vectors from ${R^0}$, ${R^1}$, …, ${R^{K-1}}$ to form a vector for each user and item:

    \begin{center}
    $\hat{y}_{eval} = u_{eval} \cdot i_{eval}$
    
    where $u_{eval} = v^0_u || … || v^{K-1}_u$
    $i_{eval} = v^0_i || … || v^{K-1}_i$
    \end{center}

These concatenated vectors are used to calculate the preference score of a user $u$ and item $i$ through inner product during evaluation.

\section{Experiments}

\subsection{Datasets and Evaluation Metrics}

To evaluate the effectiveness and scalability of our proposed method, we conduct experiments on three publicly available real-world datasets: Gowalla, Amazon-book and Yelp2018, each containing over one million implicit interactions. For each dataset, we only keep users and items with at least 10 interactions to assure a basic amount of information for each entity. The statistics of preprocessed data are summarized in Table \ref{tab:dataset}.

For all three datasets, we split 80 \% of historical interacted items of each user as the training set and the remaining as the testing set. From the training set of each dataset, we randomly pick 12.5 \% of interacted items of each user to constitute the validation set for hyper-parameters tuning. In addition, each observed user-item interaction is treated as a positive pair. In every training iteration, we conduct negative sampling to pair it with one negative item that the user has not interacted with.

For each user in the testing set, we calculate its preference scores toward all items, except for those it interacted with in the training set. We retrieve the N items with the highest preference scores as recommended ones for this user. To precisely quantify the performance of recommendation, we adopt Recall@20 and NDCG@20 for evaluation, considering both the coverage and ranking quality. Notice that we follow the implementation of Recall and NDCG in \cite{NGCF9}.

To demonstrate the advantage of our approach, we select the following five methods as our baseline models: Factorization-based method \textbf{MF-BPR} \cite{BPR}, hybrid of graph-based and model-based approach \textbf{Hop-Rec} \cite{Yang8}, and three GCN-based methods \textbf{GC-MC} \cite{Kipf10}, \textbf{PinSage} \cite{PinSage11}, and \textbf{NGCF} \cite{NGCF9}. For GC-MC and PinSage which are not originally proposed to solve the one-class recommendation problem, we keep the strategies in their proposed graph convolutional layers and use the BPR loss to optimize learned representations of entities.

\subsection{Parameter Settings}

We optimize all models except Hop-Rec with Adam optimizer and initialize parameters of all the models with Xavier. For all methods in our experiments, dimensions of user and item vectors are fixed to 256. We use inner product of user and item representations as their preference score. We employ grid search to search for the best combination of multiple hyper-parameters, including learning rate in \{5e-4, 1e-4\}, coefficient of L2 regularization in \{1e-6, 5e-7\}, and embedding dropout ratio in \{0.1, 0.2\}, respectively. As for our model, we additionally searched for the task number $K$ in the range \{2, 3, 4\}.

\subsection{Quantitative Analysis}

Table \ref{tab:perf} reports the performance of our proposed model with other five baseline methods. To begin with, MF-BPR performs the worst among all methods on all three datasets since it considers only direct (1st order) neighboring relations. HOP-Rec, despite having the same model structure as MF-BPR, exploits high-order neighbors to expand the interaction data and achieves remarkable improvements. As for GCN-based methods, PinSage and GC-MC have similar outcome for all datasets. As for NGCF, which unifies information from different orders of relationship in the bipartite graph, it achieves outstanding performance on Gowalla and Amazon-book. 

Our approach yields superior performance on all datasets. Specifically, it improves over the strongest baseline method by 5.65\%, 9.88\%, 6.42\% w.r.t. Recall and 2.34\%, 6.83\%, 4.21\% w.r.t. NDCG on Gowalla, Amazon-book, Yelp2018, respectively. This indicates that our proposed multi-tasking framework considering various preference intensities improves the generality and expressiveness of learned representation and thus leads to a better recommendation quality.

To further examine the effectiveness of the proposed multi-tasking framework and attentive graph convolutional layer, we employ our multi-tasking framework to different variants of GCN-based baseline models: GC-MC, PinSage and NGCF. Table \ref{tab:att} illustrates the experimental results on the three datasets. Note that the hyper-parameter $K$ is set as 2 for all methods. In particular, take GC-MC$_{multi-task}$ as an example, representations in each layer of GC-MC are required to satisfy their own corresponding BPR loss while the graph convolutional transformation remains the same as its original paper. Comparing numbers in Table \ref{tab:att} with Table \ref{tab:perf}, we notice that three GCN-based methods achieve remarkable improvements when they employ our multi-tasking framework. Such improvements show the effectiveness of taking various preference intensities into consideration. Moreover, table \ref{tab:att} also shows that our proposed method outperforms the other three methods even after they employ the multi-tasking framework. It indicates the effectiveness of the attentive graph convolutional layer exploited in our approach.

Table \ref{tab:taskL} reports the results of varying the task number $K$ in \{1, 2, 3, 4\}. As shown in table \ref{tab:taskL}, models with $K=2, 3, 4$ all perform better than the model with $K=1$, showing the effectiveness of considering different preference intensities. However, while higher-order information is considered as K increases, additional noises might be included as well, resulting in the inferior performances of models with $K=3, 4$ in comparison to that with $K=2$.

\section{Conclusion and Future Work}

In this work, we introduce a multi-tasking framework considering various preference intensities to improve the robustness and expressiveness of learned representations. Experiments show that our approach consistently outperforms existing methods on three large-scale datasets. We also conduct supporting experiments to examine that consideration of preference intensities and attention mechanism on graph convolutional layer are both effective. In the future, we will involve more flexible mechanisms to assign weights to each sub-task to reflect their importance instead of treating them equally. A user study will also be conducted to further assess the effectiveness of the neighbor attention we learn.

\bibliographystyle{ACM-Reference-Format}
\bibliography{sample-base}

\end{document}